\documentclass{article}

\usepackage[dblblindworkshop, final]{neurips_2025}

\workshoptitle{AI for Science Workshop}

\usepackage[utf8]{inputenc} 
\usepackage[T1]{fontenc}    
\usepackage{hyperref}       
\usepackage{url}            
\usepackage{booktabs}       
\usepackage{amsfonts}       
\usepackage{nicefrac}       
\usepackage{microtype}      
\usepackage{xcolor}         

\usepackage{algorithm, algpseudocode}
\usepackage{amsmath, amssymb}
\usepackage{graphicx}

\usepackage{listings}  
\definecolor{codegreen}{rgb}{0,0.6,0}
\definecolor{codegray}{rgb}{0.5,0.5,0.5}
\definecolor{codepurple}{rgb}{0.58,0,0.82}
\definecolor{backcolour}{rgb}{0.98,0.98,0.95}
\lstdefinestyle{mystyle}{
    backgroundcolor=\color{backcolour},   
    commentstyle=\color{codegreen},
    keywordstyle=\color{magenta},
    numberstyle=\tiny\color{codegray},
    stringstyle=\color{codepurple},
    basicstyle=\ttfamily\footnotesize,
    breakatwhitespace=false,         
    breaklines=true,                 
    captionpos=b,                    
    keepspaces=true,                 
    numbers=left,                    
    numbersep=5pt,                  
    showspaces=false,                
    showstringspaces=false,
    showtabs=false,                  
    tabsize=2
}
\usepackage{pifont}
\usepackage{subcaption}
\newcommand{\cmark}{\ding{51}}%
\newcommand{\xmark}{\ding{55}}%
\DeclareMathAlphabet{\mathbfsf}{\encodingdefault}{\sfdefault}{bx}{n}

\title{TorchQuantumDistributed}

\author{%
Oliver Knitter \quad Jonathan Mei \quad Masako Yamada \quad Martin Roetteler \\
IonQ, Applications Team \\
College Park, MD 20740 \\
\texttt{\{oliver.knitter,jmei,yamada,martin.roetteler\}@ionq.co}
}

\begin{document}

\maketitle

\begin{abstract}
  TorchQuantumDistributed (\texttt{tqd})\footnote{The code repository for \texttt{tqd} is available here: \href{https://github.com/ionq/torchquantum-dist/tree/main}{https://github.com/ionq/torchquantum-dist/tree/main}} is a PyTorch-based \citep{paszke2019pytorch} library for accelerator-agnostic differentiable quantum state vector simulation \textit{at scale}. This enables studying the behavior of learnable parameterized near-term and fault-tolerant quantum circuits with high qubit counts.
\end{abstract}

\begin{table}[!hb]
    \centering
    \begin{tabular}{c|c|c|c|c}
         & \begin{tabular}{@{}c@{}}Quantum \\ Simulation\end{tabular} & Differentiable & \begin{tabular}{@{}c@{}}Accelerator \\ Agnostic\end{tabular} & Scalable \\
         \hline
         \texttt{torch} & \xmark & \cmark & \cmark & \cmark \\
         \texttt{qiskit} & \cmark & \xmark & \xmark & \xmark \\
         \texttt{cirq} & \cmark & \xmark & \xmark & \cmark \\
         \texttt{tensorflowquantum} & \cmark & \cmark & \xmark & \cmark \\
         \texttt{pennylane} & \cmark & \cmark & --- & --- \\
         \texttt{torchquantum} & \cmark & \cmark & \cmark & \xmark \\
         \hline
         \texttt{tqd} & \cmark & \cmark & \cmark & \cmark \\
    \end{tabular}
    \caption{Features of different frameworks for quantum simulation and differentiable programming. A checkmark (\cmark) indicates full or near-full support; a cross (\xmark) indicates lacking support; and a dash (---) indicates partial support for some features individually but not together in all combinations.}
    \label{tab:features}
\end{table}

\section{Introduction}

With the increased interest in quantum computing in recent years, there has been an explosion of quantum circuit simulation tools, particularly those that facilitate quantum machine
learning. Many popular frameworks such as Qiskit \citep{javadi-abhari_quantum_2024}, Cirq \citep{developers_cirq_2025}, Pennylane \citep{bergholm_pennylane_2022}, and
TorchQuantum \citep{wang2022quantumnas}, offer quantum simulation that can be incorporated into existing Python data processing workflows. However, when using state vector simulation, these frameworks do not enable distribution of the state vector across multiple accelerators, limiting scalability and thus the use of these frameworks for studying circuits large enough to potentially yield an advantage over purely classical methods. Furthermore, many of these frameworks use software backends that are largely locked into one particular hardware ecosystem.

Leveraging the power, scaling, and hardware extensibility of PyTorch \citep{paszke2019pytorch}, this paper introduces TorchQuantumDistributed, a differentiable quantum simulation library that
enables the study of larger quantum machine learning models using distributed state vector simulator backends.

\section{Background}
\label{sec:background}

\subsection{Development of computing for subfields}

\paragraph{Quantum Computing}
Among the first notable open-source python-based frameworks for quantum simulation was ProjectQ \citep{steiger2018projectq}. Later entrants were developed by larger institutions for their own roadmaps and ecosystems: Qiskit \citep{javadi-abhari_quantum_2024} and Cirq \citep{developers_cirq_2025} provided python interfaces for defining gate-based circuits and features specific to each of their respective hardware platforms, while Q\# \citep{svore2018q} defined its own new language for quantum execution and algorithmic development. Cirq has hardware acceleration capabilities, but they are largely tied to the NVIDIA CUDA ecosystem \citep{bayraktar2023cuquantum}.

\paragraph{Deep learning}
Deep learning libraries started with the enablement of automatic differentiation on static graphs, as in Theano \citep{bergstra2011theano}, LuaTorch \citep{collobert2002torch}, and Caffe \citep{jia2014caffe}. Again, big companies released their own frameworks that have since captured a majority of the share of open-source development\footnote{as measured by \url{paperswithcode.com} implementations up to Oct 2024}: TensorFlow \citep{abadi2016tensorflow} featured strong support for deployment, while PyTorch \citep{paszke2019pytorch} gained popularity due to its extensibility and eager execution creating dynamic computation graphs, which TensorFlow eventually adopted as well. 

\paragraph{Quantum Machine Learning}
TensorFlow Quantum \citep{broughton2020tensorflow} interfaces Cirq and TensorFlow, inheriting the capabilities and drawbacks of each. TorchQuantum \citep{wang2022quantumnas} is a native PyTorch implementation of statevector simulation, so it carries some overheads but can support any hardware that PyTorch does, and it does not utilize `torch.distributed` to allow scaling across multiple accelerators. Pennylane \citep{bergholm_pennylane_2022} also provides a platform for statevector simulation supporting several different computational backend libraries and even hardware environments, placing it closest to our desired level of functionality, but ultimately also lacks the ability to distribute the state vector across multiple accelerators when using non-CUDA devices \citep{asadi2024hybrid}.

\subsection{Mathematical preliminaries}

We assume a baseline familiarity with linear algebra, probability, and statistics, and the terminology of ML, but not necessarily knowledge of quantum computing. Thus, we provide an overview of relevant quantum computing concepts.

\paragraph{Notation} Since we deal with a variety of mathematical objects, notation is treated with care. We use serif to denote a deterministic value (e.g. $x$), sans-serif to denote a random variable (e.g. $\mathsf{x}$), bold lower case to denote a (column) vector (e.g. $\mathbf{x}$ for deterministic vectors or $\mathbfsf{x}$ for random vectors), bold capital to denote matrices or tensors (e.g. $\mathbf{X}$), and blackboard to denote sets (e.g. $\mathbb{R}$). Calligraphic is used for other objects not previously listed, like distributions or operators, and their use will be clarified depending on the context (e.g. $\mathcal{M}$ as a multinomial distribution or $\mathcal{G}$ as a quantum gate). We also use teletype to refer to specific code libraries or functions (e.g. \texttt{torch.permute}).
We denote the $m$-th canonical coordinate basis for an $n$-dimensional vector space $\mathbf{e}_m^{(n)}$ as all $0$'s except a $1$ in the $m$-th coordinate, and $\mathbf{I}$ as the identity matrix.
To define the operations we use, let $\mathbf{W}$, $\mathbf{Y}$, and $\mathbf{Z}$ be tensors: $\mathbf{x}^*$ denotes the conjugate transpose of $\mathbf{x}$; $\otimes$ is an outer product; given a matrix $\mathbf{Y}$, we take $\mathbf{W}=\mathbf{Y}\times_i \mathbf{Z}$ to denote the contraction of dimension index $i$ of tensor $\mathbf{Z}$, or more explicitly, $\mathbf{W}_{\ldots j \ldots}=\sum_{\ell}\mathbf{Y}_{j\ell}\mathbf{Z}_{\ldots\ell\ldots}$, where only the $i$-th index is shown for both $\mathbf{W}$ and $\mathbf{Z}$. This can be generalized to higher dimensional tensors $\mathbf{Y}$ and ordered index sets, where for $\mathbf{W}=\mathbf{Y}\times_{\mathbb{I}} \mathbf{Z}$ we have $\mathbf{W}_{\ldots \mathbb{J} \ldots}=\sum_{\mathbb{L}}\mathbf{Y}_{\mathbb{J}\mathbb{L}}\mathbf{Z}_{\ldots\mathbb{L}\ldots}$.
While the Dirac ``bra-ket'' notation commonly used in quantum computing could be applied here, we aim to only use notation more familiar to an ML audience.

\subsubsection{Quantum computing basics}

Qubits are the basic unit of computation in quantum computing. The state of a qubit $\boldsymbol{\psi}$ can be represented as a complex vector, $\boldsymbol{\psi}\in\mathbb{C}^2$. The basis vectors $\mathbf{e}_0^{(2)}$ and $\mathbf{e}_1^{(2)}$ are known as the computational basis for a single qubit, and naturally the state can be represented in this basis as $\boldsymbol{\psi}=\sum_{i=0}^{1} \alpha_i \mathbf{e}_i^{(2)}$ for some $\alpha_i\in\mathbb{C}$. There is a constraint that $\|\boldsymbol{\alpha}\|_2=1$, and we will return to this later. A $q$-qubit state is represented in the tensor product space $\bigotimes_{i=0}^{q-1} \mathbb{C}^2 \cong \mathbb{C}^{2^q}$.

Quantum measurement is famously stochastic. 
Letting $\boldsymbol{\phi}$ be an eigenvector of a given measurement operator, then
a system in the state $\boldsymbol{\psi}$ is observed to be in the state $\boldsymbol{\phi}$ with probability given by the square magnitude of their scalar product $|\boldsymbol{\phi}^{*}\boldsymbol{\psi}|^2$. In many algorithms, the Pauli-$Z$ operator, which can be described by the matrix $\mathbf{Z}=\begin{pmatrix}
    \mathbf{e}_0^{(2)} & -\mathbf{e}_1^{(2)}
\end{pmatrix}$, is a natural choice for measurement as it is diagonalized by the computational basis. In Dirac notation, we may refer to $\mathbf{e}_0^{(2)}$ as being state "$|0\rangle$" and $\mathbf{e}_1^{(2)}$ as being state "$|1\rangle$". Returning to the constraint on $\boldsymbol{\alpha}$, we can examine the behavior of measurements of $\boldsymbol{\psi}=\sum_{i=0}^{1} \alpha_i \mathbf{e}_i^{(2)}$ in the computational basis. The probability of measuring state $|0\rangle$ is given by $|(\mathbf{e}_0^{(2)})^* \, \boldsymbol{\psi}|^2 = |(\mathbf{e}_0^{(2)})^* \, \sum_{i=0}^{1} \alpha_i \mathbf{e}_i^{(2)}|^2=|
\alpha_0|^2$. Similarly, the probability of measuring state $|1\rangle$ is $|(\mathbf{e}_1^{(2)})^* \, \boldsymbol{\psi}|^2 = |
\alpha_1|^2$. Thus we see that the constraint $|
\alpha_0|^2+|\alpha_1|^2=1$ is required to satisfy the laws of probability and describe a valid distribution.

A quantum computation can be specified in the form of a ``circuit'' consisting of a register of qubits, a sequence of unitary gate operations acting on the register, and some measurements. A comprehensive treatment of the mathematics of quantum computation can be found in \citep{nielsen2010quantum}.

\subsubsection{QML primer}
The field of QML describes the use of quantum computing to perform ML tasks. Early work focused on using quantum primitives to implement subroutines commonly found in traditional ML algorithms, but the field has also evolved to include full models and even optimization algorithms that can be used to replace large portions of the classical computing workload. A brief survey of the field is useful to provide context for our work; a more complete review of the QML field can be found in \citep{schuld2021machine}. 

Initially, QML evolved from a start in custom-designed models and methods for specific tasks~\citep{guta_quantum_2010, schuld_quantum_2014-1}.
Then, the variational eigensolver~\citep{peruzzo_variational_2014} from chemistry appeared alongside analytical formulas for estimating gradients~\citep{li_hybrid_2017} of parameterized quantum circuits, giving rise to general-purpose algorithms for which learning could be handled by the same optimization methods used to train classical neural networks ~\citep{krizhevsky_imagenet_2012}.

Another related field can be neatly summarized as, ``quantum-inspired ML,'' which borrows concepts and tools from quantum mechanics and computing, for the purpose of running algorithms on classical hardware. The methods have been applied successfully to classic ML problems such as learning time series models on graphs \citep{mei_signal_2017} and image classification \citep{stoudenmire_supervised_2016}, to further push the boundaries on quantum chemistry \citep{knitter2025retentive}, and even in conjunction with QML on actual quantum hardware for text classification \citep{kim2025quantum}. While it may not yet be clear whether quantum-inspired ML should be considered part of QML or exist as a separate area, our \texttt{tqd} framework provides a toolbox for developing both.

\section{Design Principles}
\label{sec:design}

TorchQuantum provides a pythonic statevector simulator. As suggested by the name, TorchQuantumDistributed borrows the nice features and usage style from TorchQuantum, while utilizing distributed tensors to shard the statevector across multiple accelerators, allowing us to scale statevector simulation to more qubits than previously possible.

\paragraph{Object Oriented and Functional} Mirroring PyTorch, \texttt{tqd} provides object oriented and functional interfaces to deep learning computation. This offers the flexibility to design model architecture and also extend the framework with custom quantum operations.

\paragraph{Extensible} As a starting point, a universal gate set using single-qubit Pauli rotation gates and the two-qubit ``CNOT'' is provided, which can in principle, given infinite numerical precision, perform any quantum operation \citep{nielsen2010quantum}. Of course, it may be more efficient to define certain other commonly encountered gates, reducing the number of computations needed to perform operations and simulate circuits. As in many other frameworks, \texttt{tqd} allows for easy definition of custom quantum gate operations, though users should take care to ensure they are unitary.

\paragraph{Modularity} This ties into extensibility. For ease of development, different functionalities are grouped together according to the aspects of the framework they affect. For example, the logic for implementing custom operators lives separately from the logic for tracking the dimension order of the statevector.

\paragraph{Code Parsimony} The usage of \texttt{tq} is very intuitive. However, the implementation of \texttt{tq} includes a separate file for the \texttt{nn.functional} version of a gate as well as for the stateful \texttt{nn.Module} version. \texttt{tqd} condenses this, defining gates only once in a central location and generating both function and object forms of gates operations in a programmatic fashion following a consistent schema, without sacrificing readability.

\section{Implementation Details}
\label{sec:implementation}
Here we describe the salient aspects of the implementation needed to enable state vector sharding. For expositional clarity, we will mostly deal with a single statevector corresponding to an implicit batch size of $1$ as the concepts generalize naturally to multiple statevectors. Where the batch dimension is relevant to the discussion, we will make explicit note of it.

\subsection{Bookkeeping}
A large part of the work required for simulating quantum computation, especially in scaling it up across many hardware nodes, is tracking how different dimensions of the statevector, corresponding to qubits in the circuit, are being represented and stored at any given intermediate point during the computation.

First, we describe how `tq` utilizes compute and data movement primitives in PyTorch. Then we detail speed improvements that `tqd` makes over `tq`. We initially ignore considerations for sharding the statevector, focusing on them in the later section.

\subsubsection{Tensor dimension order}
\label{sec:sub:sub:tensor_dim_order}

To describe the computation, we introduce some more notation. 
Let $\mathbf{X}\in\mathbb{C}^{\overbrace{2\times 2\times \cdots \times 2}^{q}}$ be the complex statevector for a $q$-qubit circuit. 
We take a gate $\mathcal{G}$ as defined by the ordered pair of 1) a unitary matrix $\mathbf{M}$; and 2) an ordered set of qubits on which it operates $\mathbb{Q}$.
A gate operating on a statevector is denoted $\mathcal{G}(\mathbf{X})$.

Suppose we are given $\mathbb{Q}$, the set of qubits on which a gate operates. Consider the operation $\textrm{MoveDim}(\mathbf{X}, \mathbb{Q})$, which moves the dimensions specified in the ordered set $\mathbb{Q}$ to the front. This can be implemented as \texttt{torch.movedim(X, Q, destination=torch.arange(len(Q))}. Note that if $\mathbb{Q}$ contains all integers up to $q$, then $\textrm{MoveDim}(\cdot, \cdot)$ is equivalent to permuting the dimensions (\texttt{torch.permute}).

Further, we define its inversion $\textrm{MoveDim}^{-1}(\mathbf{X}, \mathbb{Q})$ to reorder dimensions from the front such that
\begin{align}
    \mathbf{X}=\textrm{MoveDim}^{-1}(\textrm{MoveDim}(\mathbf{X}, \mathbb{Q}), \mathbb{Q}).
\end{align}

To perform the computation of the gate on the statevector $\mathbf{X}$, \texttt{tq} rearranges the dimensions, performs the computation using a (broadcasting) matrix multiplication primitive (denoted ``mm''), and then rearranges the dimensions back. This is summarized in Algorithm \ref{alg:tq_dim_order}.

\begin{algorithm}[t]
    \caption{
        \label{alg:tq_dim_order} \texttt{tq} computation pseudocode
    }
    \begin{algorithmic}
        \Require $\mathbf{X}$, $(\mathbf{M}, \mathbb{Q})$
        \Ensure $Y=\mathcal{G}(X)$
        \State $\mathbf{X} \gets$ MoveDim$(\mathbf{X}, \mathbb{Q})$
        \State $\mathbf{X}\gets \textrm{mm}(\mathbf{M}, \mathbf{X})$
        \State $\mathbf{Y} \gets$ MoveDim$^{-1}(\mathbf{X}, \mathbb{Q})$
    \end{algorithmic}
\end{algorithm}

The computation of \texttt{tqd} follows Algorithm \ref{alg:tqd_dim_order}.

\begin{algorithm}[!ht]
    \caption{
        \label{alg:tqd_dim_order} \texttt{tqd} computation pseudocode
    }
    \begin{algorithmic}
        \Require $\mathbf{X}$, $\mathbb{I}$, $(\mathbf{M}, \mathbb{Q})$
        \Ensure $\textrm{MoveDim}^{-1}(\mathbf{Y},\mathbb{J}) = \mathcal{G}(\textrm{MoveDim}^{-1}(\mathbf{X}, \mathbb{I}))$
        \State $\mathbf{X} \gets$ MoveDim$(\mathbf{X}, \mathbb{Q})$
        \State $\mathbb{J}\gets [\mathbb{Q},\;\mathbb{I}\backslash\mathbb{Q}]$  \Comment{List concatenation} 
        \State $\mathbf{Y}\gets \textrm{mm}(\mathbf{M}, \mathbf{X})$
    \end{algorithmic}
\end{algorithm}

While if implemented using \texttt{torch.permute} there is no additional data movement overhead from the additional $\textrm{MoveDim}$, it is nonetheless helpful for implementing data sharding to track the actual tensor layout in memory.

\subsection{Sharding and dimension grouping}
\begin{figure}[!ht]
    \centering
    \includegraphics[width=0.8\linewidth]{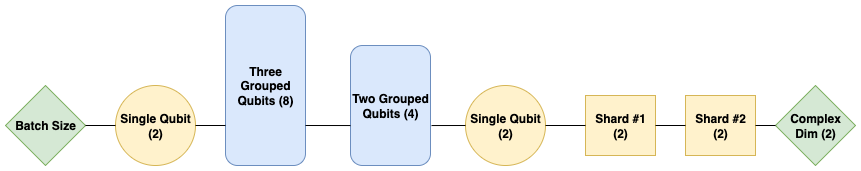}
    \caption{An example dimensional arrangement for a TQD distributed tensor representating a nine qubit statevector, with two sharded qubits. TQD always reserves the first and last dimension for batching and to contain the real and imaginary parts. Sharded qubits correspond with the dimensions preceding the final one, and at least two unsharded qubits are always kept ungrouped.}
    \label{fig:qubit_grouping}
\end{figure}
While section \ref{sec:sub:sub:tensor_dim_order} provides a high level mathematical view of the computation, there are further details to track when distributing the statevector across multiple accelerator devices. In particular, while PyTorch places no intrinsic limitation on the number of dimensions a tensor can have, it was not intended to work with arbitrary tensor ranks, and certain fundamental subroutines within the underlying architecture may place their own dimension limitations. Therefore, unlike TorchQuantum, which shapes the statevector so that each tensor dimension encodes a single qubit, \texttt{tqd} arranges statevector DTensors so that some dimensions correspond with groups of multiple qubits.

Figure \ref{fig:qubit_grouping} illustrates how \texttt{tqd} arranges a 9-qubit statevector, with 2 sharded dimensions, and a total tensor rank of 8. In practice, we find our implementation works with tensor ranks as high as 16. The first and final dimensions are always reserved for batching and for the real and imaginary parts of the statevector. Distributing across $d$ accelerators allows for sharding $\log_2d$ dimensions, each corresponding with 1 qubit. From there, \texttt{tqd} assumes at least two unsharded dimensions correspond with single qubits, and that either all qubits are ungrouped, or there exist at least two grouped qubit dimensions.

This arrangement corresponds with a linear permutation of the qubits, where continuous subsets of qubits can be grouped together. A second tensor keeps track of exactly how the qubits are arranged within this permutation. Reshaping the tensor corresponds with regrouping qubits without changing their order, while permuting the tensor dimensions corresponds with rearranging groupings of tensors in the permutation. Since reshaping and permuting tensors stored in contiguous memory is not computationally expensive, \texttt{tqd} is built on a framework of interchanging qubit positions within the grouping permutation. This allows us to easily retrieve the qubit dimensions needed to perform gate multiplications and redistribute across devices.

\subsection{Shot noise}
Shot noise describes the randomness inherent to the process of quantum measurement. Specifically, each run of the circuit ending in measurement yields a categorical variable. Typically, we run many shots of the circuit, and the many independent realizations of the categorical variable can be summarized by a multinomial variable. Because this is inherent to the way we retrieve useful information from the quantum device, we view shot noise as an essential component of quantum simulation. We describe two different ways we incorporate this stochasticity into our framework along with their benefits and drawbacks.

We can perform sampling from the exact multinomial distribution. This is slow and not differentiable. Thus, this may be preferred during inference to get more accurate characterization of the models.

In the high-shot limit, the multinomial variable can be well-approximated by a degenerate multivariate Gaussian variable \citep{georgii2012stochastics}. While not exact, this is fast and differentiable, as it is an application of the reparameterization trick \citep{kingma2014auto-encoding}. Thus, this may be more useful during training, where we still wish to run the model in a robust, noise-aware manner \citep{kim2025quantum}, but also value optimization iteration speed.

\begin{figure}[!ht]
    \includegraphics[width=0.525\columnwidth]{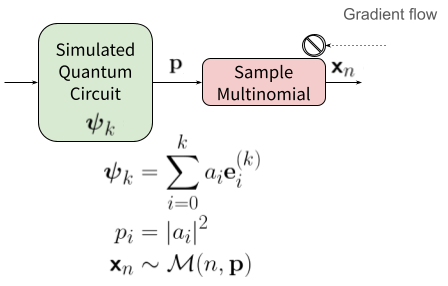}%
    \includegraphics[width=0.4\columnwidth]{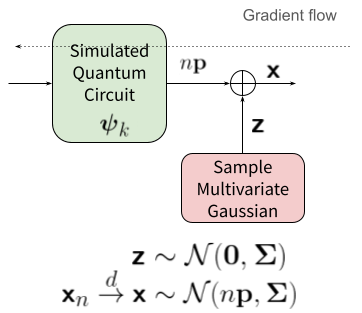}
    \caption{\label{fig:gradient_flow}Left: exact sampling breaks differentiability. Right: approximate sampling uses the reparameterization trick to maintain differentiability.}
\end{figure}

Figure \ref{fig:gradient_flow} shows the breaking of the computational graph caused by exact sampling, which is a non-differentiable operation (left), and the application of the reparameterization trick to approximate sampling from the quantum circuit in a differentiable way (right).

\subsubsection{Exact}
Sampling from a multinomial can be accomplished on a single accelerator in a straightforward call to \texttt{torch.multinomial()}.
However, when the statevector is split across multiple accelerators, the same mathematical operation is not so simple to implement faithfully. Interestingly, we can utilize several statistical properties of the multinomial to perform this in a distributed fashion.

Take $\mathbfsf{x}\sim\mathcal{M}(n, \mathbf{p})$ to be jointly multinomial variables, for parameters $0 \le p_i \le 1$ that satisfy $\sum_{i=0}^{k-1}p_i = 1$ and integer $n$. That is, $\{\mathsf{x}_i\}$ has the PMF
\begin{align}
    P(\mathsf{x}_0=x_0,\mathsf{x}_1=x_1,\ldots,\mathsf{x}_{k-1}=x_{k-1}) = p(x_0,x_1,\ldots,x_{k-1}) = \frac{n!}{\prod_{i=0}^{k-1}x_i !} \prod_{i=0}^{k-1}p_i^{x_i}
\end{align}
supported on integers $x_i$ such that $\sum_{i=0}^{k-1}x_i = n$.

Suppose we have a set of $m$ index sets $\{\mathbb{I}_i\}$ such that $\bigcup\limits_{i=0}^{m-1} \mathbb{I}_i = \{0, \ldots, k-1\}$ and $\mathbb{I}_i \bigcap \mathbb{I}_j = \varnothing \; \forall i \ne j$. That is, the index set is non-overlapping and covers the integer indices up to but excluding $k$. Then consider the variables $\mathsf{y}_j = \sum_{i\in\mathbb{I}_j}\mathsf{x}_i$. Then the variables $\mathsf{y}_j$ are themselves multinomial with parameters $q_j=\sum_{i\in\mathbb{I}_j}p_i$ \citep{georgii2012stochastics},
\begin{align}
    \mathbfsf{y} \sim \mathcal{M}(n, \mathbf{q}).
\end{align}
Next, we note that the variables $\{\mathsf{x}_i : i \in \mathbb{I}_j\}|\mathsf{y}_j$ are also multinomial \cite{georgii2012stochastics}. This points to a hierarchical sampling scheme, which groups individual local variables $\mathsf{x}_i$ to form intermediate $\mathsf{y}_j$, then samples values of $\mathsf{y}_j$, and finally samples the $\mathsf{x}_i$'s conditional upon the $\mathsf{y}_j$'s. Practically, we can associate each accelerator with its own $\mathbb{I}_j$ with the indices for variables for which it locally contains the parameters. One small round of initial communication is required where each accelerator shares its $q_j$ so that they can all can assemble the full vector $\mathbf{q}$. With a shared seed, each accelerator can then sample $\mathbfsf{y}$ in parallel, then taking $\mathsf{y}_j$ to sample $\mathsf{x}_i$.

\subsubsection{Approximate}

Let the diagonal matrix $\mathbf{D}$ have diagonal elements $\mathbf{D}_{ii}=\sqrt{p_i}$. As the number of shots $n$ grows, we have a certain convergence in distribution to a multivariate Gaussian

\begin{align}
   \mathbfsf{y}_n &\sim \mathcal{M}(\mathbf{p}, n) \\
   \mathbfsf{x}_n &= \frac{1}{\sqrt{n}}\mathbf{D}^{-1}(\mathbfsf{y}_n - n\mathbf{p}) \\
   \mathbfsf{x} &\sim \mathcal{N}(0, \boldsymbol{\Sigma})\\
   \Rightarrow \mathbfsf{x}_n &\overset{d}{\rightarrow} \; \mathbfsf{x}
\end{align}
where $\boldsymbol{\Sigma}_{ii}=p_i(1-p_i)$ and $\boldsymbol{\Sigma}_{ij}=-p_i p_j$ for $i\ne j$ \citep{georgii2012stochastics}. One way to sample from the Gaussian is to have access to $\mathbf{S}$, a matrix factorization of the covariance satisfying $\boldsymbol{\Sigma}=\mathbf{S}\mathbf{S}^\top$. Then we can undo the transformation from $\mathbfsf{x}_n$ to $\mathbfsf{y}_n$ to approximate the multinomial sample. We use a factorization based on a Householder transformation that can be fairly easily computed. Let the unnormalized vector $\widetilde{\mathbf{v}}=\mathbf{e}_k^{(k)} - \mathbf{u}$ and its normalization $\mathbf{v}=\widetilde{\mathbf{v}}/\|\widetilde{\mathbf{v}}\|_2$ (in the degenerate case where $\mathbf{u}=\mathbf{e}_k^{(k)}$, we take $\mathbf{v}=\mathbf{0}$). Then we can take 
\begin{align}
    \mathbf{S} = \mathbf{I} - \mathbf{v}\mathbf{v}^\top.
\end{align}
Finally, for completeness, taking i.i.d. unit normals $\mathsf{z}_i\sim\mathcal{N}(0, 1)$ for $i=0,\ldots,k-2$ and $\mathbfsf{z}=\begin{pmatrix}
    \mathsf{z}_0 & \ldots & \mathsf{z}_{k-2} & 0
\end{pmatrix}^\top$, we have
\begin{align}
    \mathbfsf{y}_n &= n\mathbf{p} + \sqrt{n}\mathbf{D}\mathbfsf{x}_n\\
    \mathbfsf{x}_n &= \mathbf{S}\mathbfsf{z}\\
    \Rightarrow \mathbfsf{y}_n &= n\mathbf{p}+\sqrt{n}\mathbf{D}\mathbf{S}\mathbfsf{z}.
\end{align}

\subsection{Backpropagating Invertible Computations}
Since quantum computing is built on unitary operations, the computations are invertible. This means that for backpropagation, we may recompute intermediate activations instead of storing them. For memory-intensive settings such as quantum simulation, this can be used to ameliorate some of the memory requirements for enabling differentiability.

Concretely, let $\partial_{\mathbf{x}}$ ($\partial_{\mathbf{U}
}$) denote the gradient of the loss function with respect to vector (matrix) variable $\mathbf{x}$ ($\mathbf{U}$). Then consider a layer implementing unitary matrix multiplication,
\begin{align}
    &\quad\,\mathbf{y} = \mathbf{U}\mathbf{x}\\
    \Rightarrow& \begin{cases}
        \partial_\mathbf{x} = \mathbf{U}^\top\partial_{\mathbf{y}}\\
        \partial_{\mathbf{U}} = \partial_{\mathbf{y}}\mathbf{x}^\top \label{eq:save_x}.
    \end{cases}
\end{align}
Generally, our automatic differentiation framework would ``see'' Equation \eqref{eq:save_x} and store the intermediate activation $\mathbf{x}$ to compute the gradient $\partial_\mathbf{U}$. However, since the inverse of a unitary is its conjugate transpose, we can instead implement the backward pass of Equation \eqref{eq:save_x} using
\begin{align}
    \mathbf{x} = \mathbf{U}^{*}\mathbf{y}.
\end{align}
This removes the need to store the layer input $\mathbf{x}$ at the expense of recomputing it from layer output $\mathbf{y}$.

\section{Profiling}
\label{sec:profiling}
\begin{figure}[hbt]
    \centering
    \begin{subfigure}[c]{0.4\linewidth}
    \centering
        \includegraphics[width=\linewidth]{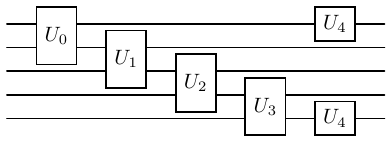}
    \end{subfigure}%
    \hspace{0.2\linewidth}
    \begin{subfigure}[c]{0.2\linewidth}
    \centering
        \includegraphics[width=\linewidth]{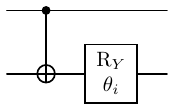}
    \end{subfigure}
    \caption{The primary building block for the ansatz (left) consists of unitaries with a ladder structure across the width of the circuit. Each block $U_i$ is defined as a combination of a controlled NOT and a single-qubit rotation $R_Y$ of angle $\theta_i$ (right).}
    \label{fig:unitary}
\end{figure}
As a fundamental proof of concept, we conduct a basic profiling test of \texttt{tqd} by simulating a rudimentary ansatz using a multi-node HPC cluster. Each node within the cluster contains 4 AMD MI250X accelerators, each containing 2 graphics compute dies, which act effectively as separate 64GB GPUs. For these profiling experiments, we use the circuit structure shown in figure \ref{fig:unitary}: a ladder structure consisting of entangling CNOT and Pauli Y rotation gates, since it is a commonly used ansatz component throughout QML models \citep{kim2025quantum}. With a batch size of 16, we run 5 iterations of Adam \citep{kingma15adam} to optimize the initial circuit input, but not the ansatz parameters. Information from the final gradient update step is recorded for profiling. More specifically, for a single GPU, the first according to the world topology, we record the total walltime, the total time dedicated to all-to-all communication, and the total memory used, though not simultaneously utilized, for all GPU operations within the final update step. These three benchmarks were chosen as qualitative measures of the leading compute costs with respect to problem size and accelerator count.

\subsection{Scaling}

\begin{figure}[hbt]
    \centering
    \includegraphics[width=\linewidth]{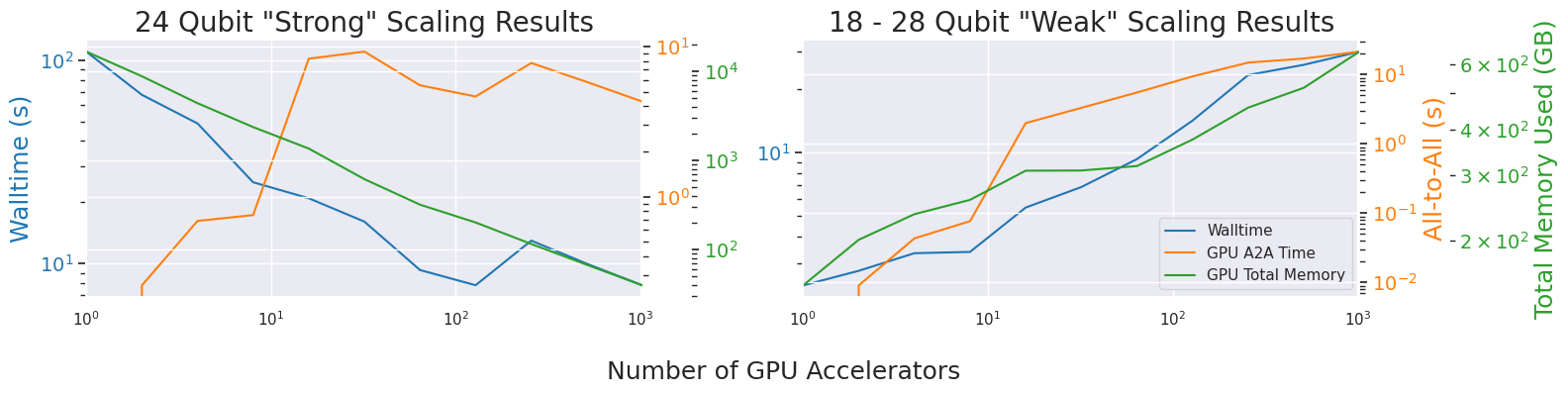}
    \caption{Basic "strong" and "weak" scaling tests of \texttt{tqd}, applying the ansatz from figure \ref{fig:unitary} to qubit sizes between 18 and 28. We vary the number of accelerators between 1 and 1024. We perform benchmarking by collecting the walltime, total NCCL all-to-all communication time, and total memory usage recorded by a single GPU for one forward--backward pass through the ansatz.}
    \label{fig:scaling}
\end{figure}

The results of these benchmarking experiments are depicted in figure \ref{fig:scaling}. Here, we perform a "strong" and "weak" scaling test as the total number of GPUs is doubled incrementally from 1 to 1024. Our use of quotations is intended to indicate that these are only basic proof of concept tests, not thorough explorations of the full extent of \texttt{tqd}'s capabilities. For the "strong" scaling test, we perform a 24-qubit test at all world sizes, while for the "weak" scaling test, we incrementally increase problem size from 18 to 28 qubits. Presented on a log--log scale, these results in both cases indicate favorable power law trends between world size and nearly all benchmarks: for the "strong" scaling test, the additional overhead incurred by inter-GPU communication does not appear to substantially hinder the expected improvement in walltime.

\section{Conclusion}
\label{sec:conclusion}

We have introduced \texttt{tqd}, a tool for scalable hardware-accelerated, differentiable quantum statevector simulation that is extensible. We have shown basic proofs of concept indicating favorable scaling behavior when applying \texttt{tqd} to circuit simulations inspired by common QML ansatze, and it is our hope that this tool may find itself incorporated into many future QML research pipelines.

Potential future work on this simulator includes profiling peak GPU usage, network traffic, and incorporation of circuit-cutting and knitting techniques to ameliorate the communication costs.

\begin{ack}
This research used resources of the Oak Ridge Leadership Computing Facility at the Oak Ridge National Laboratory, which is supported by the Office of Science of the U.S. Department of Energy under Contract No. DE-AC05-00OR22725.
\end{ack}

\bibliographystyle{plainnat}
\bibliography{neurips}  

\begin{thebibliography}{28}
\providecommand{\natexlab}[1]{#1}
\providecommand{\url}[1]{\texttt{#1}}
\expandafter\ifx\csname urlstyle\endcsname\relax
  \providecommand{\doi}[1]{doi: #1}\else
  \providecommand{\doi}{doi: \begingroup \urlstyle{rm}\Url}\fi

\bibitem[Abadi et~al.(2016)Abadi, Barham, Chen, Chen, Davis, Dean, Devin, Ghemawat, Irving, Isard, et~al.]{abadi2016tensorflow}
Mart{\'\i}n Abadi, Paul Barham, Jianmin Chen, Zhifeng Chen, Andy Davis, Jeffrey Dean, Matthieu Devin, Sanjay Ghemawat, Geoffrey Irving, Michael Isard, et~al.
\newblock {TensorFlow}: a system for {Large-Scale} machine learning.
\newblock In \emph{12th USENIX symposium on operating systems design and implementation (OSDI 16)}, pages 265--283, 2016.

\bibitem[Asadi et~al.(2024)Asadi, Dusko, Park, Michaud-Rioux, Schoch, Shu, Vincent, and O'Riordan]{asadi2024hybrid}
Ali Asadi, Amintor Dusko, Chae-Yeun Park, Vincent Michaud-Rioux, Isidor Schoch, Shuli Shu, Trevor Vincent, and Lee~James O'Riordan.
\newblock {Hybrid quantum programming with Pennylane Lightning on HPC platforms}.
\newblock \emph{arXiv preprint arXiv:2403.02512}, 2024.

\bibitem[Bayraktar et~al.(2023)Bayraktar, Charara, Clark, Cohen, Costa, Fang, Gao, Guan, Gunnels, Haidar, et~al.]{bayraktar2023cuquantum}
Harun Bayraktar, Ali Charara, David Clark, Saul Cohen, Timothy Costa, Yao-Lung~L Fang, Yang Gao, Jack Guan, John Gunnels, Azzam Haidar, et~al.
\newblock {cuQuantum SDK}: A high-performance library for accelerating quantum science.
\newblock In \emph{2023 IEEE International Conference on Quantum Computing and Engineering (QCE)}, volume~1, pages 1050--1061. IEEE, 2023.

\bibitem[Bergholm et~al.(2022)Bergholm, Izaac, Schuld, Gogolin, Ahmed, Ajith, Alam, Alonso-Linaje, AkashNarayanan, Asadi, Arrazola, Azad, Banning, Blank, Bromley, Cordier, Ceroni, Delgado, Matteo, Dusko, Garg, Guala, Hayes, Hill, Ijaz, Isacsson, Ittah, Jahangiri, Jain, Jiang, Khandelwal, Kottmann, Lang, Lee, Loke, Lowe, McKiernan, Meyer, Montañez-Barrera, Moyard, Niu, O'Riordan, Oud, Panigrahi, Park, Polatajko, Quesada, Roberts, Sá, Schoch, Shi, Shu, Sim, Singh, Strandberg, Soni, Száva, Thabet, Vargas-Hernández, Vincent, Vitucci, Weber, Wierichs, Wiersema, Willmann, Wong, Zhang, and Killoran]{bergholm_pennylane_2022}
Ville Bergholm, Josh Izaac, Maria Schuld, Christian Gogolin, Shahnawaz Ahmed, Vishnu Ajith, M.~Sohaib Alam, Guillermo Alonso-Linaje, B.~AkashNarayanan, Ali Asadi, Juan~Miguel Arrazola, Utkarsh Azad, Sam Banning, Carsten Blank, Thomas~R. Bromley, Benjamin~A. Cordier, Jack Ceroni, Alain Delgado, Olivia~Di Matteo, Amintor Dusko, Tanya Garg, Diego Guala, Anthony Hayes, Ryan Hill, Aroosa Ijaz, Theodor Isacsson, David Ittah, Soran Jahangiri, Prateek Jain, Edward Jiang, Ankit Khandelwal, Korbinian Kottmann, Robert~A. Lang, Christina Lee, Thomas Loke, Angus Lowe, Keri McKiernan, Johannes~Jakob Meyer, J.~A. Montañez-Barrera, Romain Moyard, Zeyue Niu, Lee~James O'Riordan, Steven Oud, Ashish Panigrahi, Chae-Yeun Park, Daniel Polatajko, Nicolás Quesada, Chase Roberts, Nahum Sá, Isidor Schoch, Borun Shi, Shuli Shu, Sukin Sim, Arshpreet Singh, Ingrid Strandberg, Jay Soni, Antal Száva, Slimane Thabet, Rodrigo~A. Vargas-Hernández, Trevor Vincent, Nicola Vitucci, Maurice Weber, David Wierichs, Roeland Wiersema, Moritz
  Willmann, Vincent Wong, Shaoming Zhang, and Nathan Killoran.
\newblock {PennyLane}: {Automatic} differentiation of hybrid quantum-classical computations, July 2022.
\newblock URL \url{http://arxiv.org/abs/1811.04968}.
\newblock arXiv:1811.04968 [quant-ph].

\bibitem[Bergstra et~al.(2011)Bergstra, Bastien, Breuleux, Lamblin, Pascanu, Delalleau, Desjardins, Warde-Farley, Goodfellow, Bergeron, et~al.]{bergstra2011theano}
James Bergstra, Fr{\'e}d{\'e}ric Bastien, Olivier Breuleux, Pascal Lamblin, Razvan Pascanu, Olivier Delalleau, Guillaume Desjardins, David Warde-Farley, Ian Goodfellow, Arnaud Bergeron, et~al.
\newblock Theano: Deep learning on {GPU}s with python.
\newblock In \emph{NIPS 2011, BigLearning Workshop, Granada, Spain}, volume~3. Citeseer, 2011.

\bibitem[Broughton et~al.(2020)Broughton, Verdon, McCourt, Martinez, Yoo, Isakov, Massey, Halavati, Niu, Zlokapa, et~al.]{broughton2020tensorflow}
Michael Broughton, Guillaume Verdon, Trevor McCourt, Antonio~J Martinez, Jae~Hyeon Yoo, Sergei~V Isakov, Philip Massey, Ramin Halavati, Murphy~Yuezhen Niu, Alexander Zlokapa, et~al.
\newblock Tensorflow quantum: A software framework for quantum machine learning.
\newblock \emph{arXiv preprint arXiv:2003.02989}, 2020.

\bibitem[Collobert et~al.(2002)Collobert, Bengio, and Mari{\'e}thoz]{collobert2002torch}
Ronan Collobert, Samy Bengio, and Johnny Mari{\'e}thoz.
\newblock Torch: a modular machine learning software library.
\newblock \emph{Technical Report IDIAP-RR 02-46, IDIAP}, 2002.

\bibitem[Developers(2025)]{developers_cirq_2025}
Cirq Developers.
\newblock \emph{Cirq}.
\newblock Zenodo, July 2025.
\newblock \doi{10.5281/ZENODO.4062499}.
\newblock URL \url{https://zenodo.org/doi/10.5281/zenodo.4062499}.

\bibitem[Georgii(2012)]{georgii2012stochastics}
Hans-Otto Georgii.
\newblock \emph{Stochastics: introduction to probability and statistics}.
\newblock Walter de Gruyter, 2012.

\bibitem[Guţă and Kotłowski(2010)]{guta_quantum_2010}
Mădălin Guţă and Wojciech Kotłowski.
\newblock Quantum learning: asymptotically optimal classification of qubit states.
\newblock \emph{New Journal of Physics}, 12\penalty0 (12):\penalty0 123032, December 2010.
\newblock ISSN 1367-2630.
\newblock \doi{10.1088/1367-2630/12/12/123032}.
\newblock URL \url{https://dx.doi.org/10.1088/1367-2630/12/12/123032}.

\bibitem[Javadi-Abhari et~al.(2024)Javadi-Abhari, Treinish, Krsulich, Wood, Lishman, Gacon, Martiel, Nation, Bishop, Cross, Johnson, and Gambetta]{javadi-abhari_quantum_2024}
Ali Javadi-Abhari, Matthew Treinish, Kevin Krsulich, Christopher~J. Wood, Jake Lishman, Julien Gacon, Simon Martiel, Paul~D. Nation, Lev~S. Bishop, Andrew~W. Cross, Blake~R. Johnson, and Jay~M. Gambetta.
\newblock Quantum computing with {Qiskit}, June 2024.
\newblock URL \url{http://arxiv.org/abs/2405.08810}.
\newblock arXiv:2405.08810 [quant-ph].

\bibitem[Jia et~al.(2014)Jia, Shelhamer, Donahue, Karayev, Long, Girshick, Guadarrama, and Darrell]{jia2014caffe}
Yangqing Jia, Evan Shelhamer, Jeff Donahue, Sergey Karayev, Jonathan Long, Ross Girshick, Sergio Guadarrama, and Trevor Darrell.
\newblock {Caffe: Convolutional Architecture for Fast Feature Embedding}.
\newblock \emph{arXiv preprint arXiv:1408.5093}, 2014.

\bibitem[Kim et~al.(2025)Kim, Mei, Girotto, Yamada, and Roetteler]{kim2025quantum}
Sang Kim, Jonathan Mei, Claudio Girotto, Masako Yamada, and Martin Roetteler.
\newblock Quantum language model fine tuning.
\newblock \emph{IEEE International Conference on Quantum Computing and Engineering}, 2025.

\bibitem[Kingma and Ba(2015)]{kingma15adam}
Diederik~P. Kingma and Jimmy Ba.
\newblock Adam: A method for stochastic optimization.
\newblock In \emph{ICLR (Poster)}, 2015.

\bibitem[Kingma and Welling(2014)]{kingma2014auto-encoding}
Diederik~P. Kingma and Max Welling.
\newblock Auto-{Encoding} {Variational} {Bayes}.
\newblock In \emph{ICLR}, Banff, AB, Canada, April 2014.
\newblock URL \url{https://openreview.net/forum?id=33X9fd2-9FyZd}.

\bibitem[Knitter et~al.(2025)Knitter, Zhao, Stokes, Ganahl, Leichenauer, and Veerapaneni]{knitter2025retentive}
Oliver Knitter, Dan Zhao, James Stokes, Martin Ganahl, Stefan Leichenauer, and Shravan Veerapaneni.
\newblock Retentive neural quantum states: efficient ans{\"a}tze for ab initio quantum chemistry.
\newblock \emph{Machine Learning: Science and Technology}, 6\penalty0 (2):\penalty0 025022, 2025.

\bibitem[Krizhevsky et~al.(2012)Krizhevsky, Sutskever, and Hinton]{krizhevsky_imagenet_2012}
Alex Krizhevsky, Ilya Sutskever, and Geoffrey~E Hinton.
\newblock {ImageNet} {Classification} with {Deep} {Convolutional} {Neural} {Networks}.
\newblock In \emph{Advances in {Neural} {Information} {Processing} {Systems}}, volume~25. Curran Associates, Inc., 2012.
\newblock URL \url{https://proceedings.neurips.cc/paper/2012/hash/c399862d3b9d6b76c8436e924a68c45b-Abstract.html}.

\bibitem[Li et~al.(2017)Li, Yang, Peng, and Sun]{li_hybrid_2017}
Jun Li, Xiaodong Yang, Xinhua Peng, and Chang-Pu Sun.
\newblock Hybrid {Quantum}-{Classical} {Approach} to {Quantum} {Optimal} {Control}.
\newblock \emph{Physical Review Letters}, 118\penalty0 (15):\penalty0 150503, April 2017.
\newblock \doi{10.1103/PhysRevLett.118.150503}.
\newblock URL \url{https://link.aps.org/doi/10.1103/PhysRevLett.118.150503}.
\newblock Publisher: American Physical Society.

\bibitem[Mei and Moura(2017)]{mei_signal_2017}
Jonathan Mei and José M.~F. Moura.
\newblock Signal {Processing} on {Graphs}: {Causal} {Modeling} of {Unstructured} {Data}.
\newblock \emph{IEEE Transactions on Signal Processing}, 65\penalty0 (8):\penalty0 2077--2092, April 2017.
\newblock ISSN 1941-0476.
\newblock \doi{10.1109/TSP.2016.2634543}.
\newblock URL \url{https://ieeexplore.ieee.org/abstract/document/7763882}.
\newblock Conference Name: IEEE Transactions on Signal Processing.

\bibitem[Nielsen and Chuang(2010)]{nielsen2010quantum}
Michael~A Nielsen and Isaac~L Chuang.
\newblock \emph{Quantum computation and quantum information}.
\newblock Cambridge university press, 2010.

\bibitem[Paszke et~al.(2019)Paszke, Gross, Massa, Lerer, Bradbury, Chanan, Killeen, Lin, Gimelshein, Antiga, et~al.]{paszke2019pytorch}
Adam Paszke, Sam Gross, Francisco Massa, Adam Lerer, James Bradbury, Gregory Chanan, Trevor Killeen, Zeming Lin, Natalia Gimelshein, Luca Antiga, et~al.
\newblock Pytorch: An imperative style, high-performance deep learning library.
\newblock \emph{Advances in neural information processing systems}, 32, 2019.

\bibitem[Peruzzo et~al.(2014)Peruzzo, McClean, Shadbolt, Yung, Zhou, Love, Aspuru-Guzik, and O’Brien]{peruzzo_variational_2014}
Alberto Peruzzo, Jarrod McClean, Peter Shadbolt, Man-Hong Yung, Xiao-Qi Zhou, Peter~J. Love, Alán Aspuru-Guzik, and Jeremy~L. O’Brien.
\newblock A variational eigenvalue solver on a photonic quantum processor.
\newblock \emph{Nature Communications}, 5\penalty0 (1):\penalty0 4213, July 2014.
\newblock ISSN 2041-1723.
\newblock \doi{10.1038/ncomms5213}.
\newblock URL \url{https://www.nature.com/articles/ncomms5213}.
\newblock Publisher: Nature Publishing Group.

\bibitem[Schuld and Petruccione(2021)]{schuld2021machine}
Maria Schuld and Francesco Petruccione.
\newblock \emph{Machine Learning with Quantum Computers}.
\newblock Springer, 2021.

\bibitem[Schuld et~al.(2014)Schuld, Sinayskiy, and Petruccione]{schuld_quantum_2014-1}
Maria Schuld, Ilya Sinayskiy, and Francesco Petruccione.
\newblock Quantum {Computing} for {Pattern} {Classification}.
\newblock In Duc-Nghia Pham and Seong-Bae Park, editors, \emph{{PRICAI} 2014: {Trends} in {Artificial} {Intelligence}}, pages 208--220, Cham, 2014. Springer International Publishing.
\newblock ISBN 978-3-319-13560-1.
\newblock \doi{10.1007/978-3-319-13560-1_17}.

\bibitem[Steiger et~al.(2018)Steiger, H{\"a}ner, and Troyer]{steiger2018projectq}
Damian~S Steiger, Thomas H{\"a}ner, and Matthias Troyer.
\newblock Projectq: an open source software framework for quantum computing.
\newblock \emph{Quantum}, 2:\penalty0 49, 2018.

\bibitem[Stoudenmire and Schwab(2016)]{stoudenmire_supervised_2016}
Edwin Stoudenmire and David~J Schwab.
\newblock Supervised {Learning} with {Tensor} {Networks}.
\newblock In \emph{Advances in {Neural} {Information} {Processing} {Systems}}, volume~29. Curran Associates, Inc., 2016.
\newblock URL \url{https://proceedings.neurips.cc/paper/2016/hash/5314b9674c86e3f9d1ba25ef9bb32895-Abstract.html}.

\bibitem[Svore et~al.(2018)Svore, Geller, Troyer, Azariah, Granade, Heim, Kliuchnikov, Mykhailova, Paz, and Roetteler]{svore2018q}
Krysta Svore, Alan Geller, Matthias Troyer, John Azariah, Christopher Granade, Bettina Heim, Vadym Kliuchnikov, Mariia Mykhailova, Andres Paz, and Martin Roetteler.
\newblock Q\# enabling scalable quantum computing and development with a high-level {DSL}.
\newblock In \emph{Proceedings of the real world domain specific languages workshop 2018}, pages 1--10, 2018.

\bibitem[Wang et~al.(2022)Wang, Ding, Gu, Li, Lin, Pan, Chong, and Han]{wang2022quantumnas}
Hanrui Wang, Yongshan Ding, Jiaqi Gu, Zirui Li, Yujun Lin, David~Z Pan, Frederic~T Chong, and Song Han.
\newblock Quantum{NAS}: Noise-adaptive search for robust quantum circuits.
\newblock In \emph{The 28th IEEE International Symposium on High-Performance Computer Architecture (HPCA-28)}, 2022.

\end{thebibliography}


\appendix

\section{API Example Usage}

We will now show some simple examples of how to use \texttt{tqd}.

First, in Listing \ref{code:basic_example.py}, we see the basic building blocks for circuit construction: initializing a quantum circuit, applying gates in 3 different paradigms, and finally taking a measurement.

\begin{lstlisting}[language=Python,caption=Basic usage example: \texttt{basic\_example.py},label=code:basic_example.py]
import torch
import tqd

nq = 6  # number of qubits
qdev = tqd.DistributedQuantumDevice(nq)

# functional on the qdev
tqd.z(qdev)

# create a stateful RY gate module that tracks its own parameters
ry = tqd.RY(wires=[0], params=torch.pi/3)
ry(qdev)

# operate directly using qdev's own methods
qdev.cx(wires=[0,1])

exact = tqd.measure_allZ(qdev)
\end{lstlisting}


This snippet could be run with 2 GPUs as
\begin{lstlisting}[language=bash]
    torchrun --nproc-per-node=2 basic_example.py
\end{lstlisting}

For a more sophisticated example, we now show the differentiability, composability, and invertibility in Listing \ref{code:main_example.py}.
\begin{lstlisting}[language=Python,caption=Typical usage: \texttt{main\_example.py},label=code:main_example.py]
import torch
import tqd
import tqd.module

rank = os.environ['RANK']

qdev = tqd.DistributedQuantumDevice(
    nq,
    bsz=batch,
    device=f'cuda',
    world_sz=world_sz,
    invertible=True,
)

func_list = [
    {'func': 'ry', 'wires': [i], 'input_idx': [i]}
    for i in range(nq)
]
enc = tqd.GeneralEncoder(func_list)
base_mod = [enc]
for _ in range(3):  # 3 sets of alternating CNOT ladders and RYs
    base_mod = base_mod + \
        [tqd.CX(wires=[i, (i+1) % nq]) for i in range(nq)] + \
        [tqd.RY(wires=[i]) for i in range(nq)]
mod = tqd.module.InvertibleUnitary(base_mod)
mod.train()

def fun(qdev, inp):
    qdev.reset_states()
    mod(qdev, inp)
    meas_approx = tqd.measure_allZ(qdev, shots=0, training=True)
    return meas_approx

# test backprop:
x_i = torch.nn.Parameter(torch.rand([batch, nq], device=f'cuda:{rank}') * torch.pi / 3)

opt = torch.optim.Adam([x_i])
loss_s = []
for i in range(10):

    opt.zero_grad()
    out_i = fun(qdev, x_i)

    loss = out_i.abs().sum()
    loss_s.append(loss.item())
    loss.backward()
    opt.step()
\end{lstlisting}


\end{document}